\documentclass[aps,prl,twocolumn,showpacs,floatfix]{revtex4}
\usepackage{graphics}% Include figure files
\usepackage{dcolumn}% Align table columns on decimal point
\usepackage{bm}% bold math

\begin{document}

\title{Tests of the Gravitational Inverse-Square Law below the Dark-Energy Length Scale}

\author{D.J. Kapner}\altaffiliation[Present address: ]{Kavli Institute for Cosmological Physics, University of Chicago, Chicago IL, 60637}
\author{T.S. Cook}
\author{E.G. Adelberger}
\author {J.H. Gundlach}
\author{B.R. Heckel}
\author{C.D. Hoyle}
\author{H.E. Swanson}
\affiliation{Center for Experimental Nuclear Physics and Astrophysics, Box 354290,
University of
Washington, Seattle, WA 98195-4290}
\date{\today}

\begin{abstract}
We conducted three torsion-balance experiments to test the gravitational inverse-square law at separations between 9.53 mm and 55 $\mu$m, probing distances less than the dark-energy length scale $\lambda_{\rm d}=\sqrt[4]{\hbar c/\rho_{\rm d}}\approx 85~\mu$m. We find with 95\% confidence that the inverse-square law holds ($|\alpha| \leq 1$) down to a length scale $\lambda = 56~\mu$m and that an extra dimension must have a size $R \leq 44~\mu$m.
\end{abstract}
\maketitle
\setlength{\floatsep}{0.25in}
Recent cosmological observations\cite{ri:98,pe:99,be:03} have shown that 
70\% of all the mass and energy of the Universe is a mysterious ``dark energy'' with a density $\rho_{\rm d} \approx 3.8$ keV/cm$^3$ and a repulsive gravitational effect. 
This dark-energy density corresponds to a distance $\lambda_{\rm d}=\sqrt[4]{\hbar c/\rho_{\rm d}}\approx 85~\mu$m that may represent a fundamental length scale of gravity\cite{be:97,dv:02}.  
Although quantum-mechanical vacuum energy should have a repulsive gravitational effect, the observed $\rho_{\rm d}$ is between $10^{60}$ to $10^{120}$ times smaller than the vacuum energy density computed according to the standard laws of quantum mechanics. Sundrum\cite{su:04} has suggested that this huge discrepancy (the ``cosmological constant problem'') could be resolved if the graviton were a ``fat'' object with a size comparable to  $\lambda_{\rm d}$ that would prevent it from ``seeing'' the short-distance physics that dominates the vacuum energy. His  
scenario implies that the gravitational force would {\em weaken} for objects separated by distances $s \alt \lambda_{\rm d}$. Dvali, Gabadaze and Senjanov\'{\i}c\cite{dv:99} argue that a similar weakening of gravity could occur if there are extra {\em time} dimensions. In their scenario, the standard model particles are localized in ``our'' time, while the gravitons propagate in the extra time dimension(s) as well. Other scenarios predict the opposite behavior: the extra {\em space} dimensions of M-theory would cause the gravitational force to get {\em stronger} for $s \alt R$ where $R$ is the size of the largest compactified dimension\cite{ar:98}. These considerations, plus others involving new forces from the exchange of proposed scalar or vector particles\cite{ad:03} motivated the tests of the gravitational inverse-square law we report in this Letter. 
%
% Fig 1
%
\begin{figure}
\hfil\scalebox{.56}{\includegraphics*[185pt,24pt][510pt,300pt]{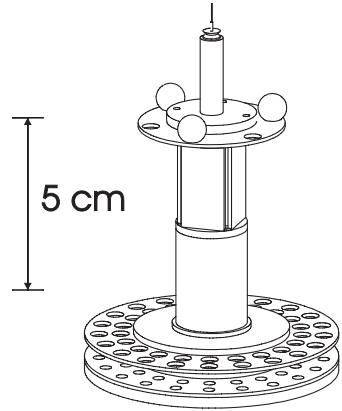}}\hfil
\caption{Scale drawing of our detector and attractor. The 3 small spheres near the top of the detector were used for a continuous gravitational calibration of the torque scale. Four rectangular plane mirrors below the spheres are part of the twist-monitoring system. The detector's electrical shield is not shown.}
\label{fig: 21 hole instrument}
\end{figure}

\begin{figure}
\hfil\scalebox{.43}{\includegraphics*[0.6in,0.4in][7.5in,5.9in]{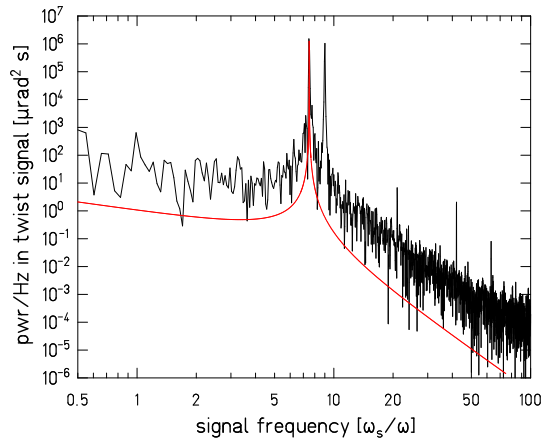}}\hfil
\caption{Fourier transform of the raw twist signal in Experiment III taken at $s=67~\mu$m (a detector-membrane separation of 46 $\mu$m). The detector's free
resonance occurs at 7.5$\omega$ and the gravitational calibration is at 9$\omega$. The peaks at 21, 42 and 63$\omega$ probe the inverse square law. The smooth curve shows the thermal noise level. At this small separation the torque noise power retains the expected $1/f$ form, but its amplitude exceeds the thermal value by about a factor of four.}
\label{fig: fft}
\end{figure}

Our tests were made with a substantially upgraded version of the ``missing mass'' torsion-balance instrument used in our previous inverse-square-law 
tests\cite{ho:01,ho:04}.
The instrument used in this work\cite{ka:05}, shown in Fig. \ref{fig: 21 hole instrument}, consisted of a torsion-pendulum detector suspended by a thin $\approx 80$-cm-long tungsten fiber above an attractor that was rotated with a uniform angular velocity $\omega$ by a geared-down stepper motor. The detector's 42 test bodies were 4.767-mm-diameter cylindrical holes machined into a 0.997-mm-thick molybdenum detector ring. The hole centers were arrayed in two circles, each of which had 21-fold azimuthal symmetry. The attractor had a similar 21-fold azimuthal symmetry and consisted of a 0.997 mm thick molybdenum disc with 42 3.178-mm-diameter holes mounted atop a thicker tantalum disc containing 21 6.352-mm-diameter
holes. The gravitational interaction between the missing masses of the detector and attractor holes applied a torque on the detector that oscillated 21 times for each revolution of the attractor, giving torques at 21$\omega$, 42$\omega$, 63$\omega$, etc. that we measured by monitoring the pendulum twist with an autocollimator system. The holes in the lower attractor ring were displaced azimuthally by 360/42 degrees and were designed to nearly cancel the 21$\omega$ torque {\em if the inverse-square law holds}. On the other hand, an interaction that violated the inverse-square law, which we
parameterize as a single Yukawa 
\begin{equation}
V(r)=-G \frac{m_1 m_2}{r} [1+\alpha \exp(-r/\lambda)]~,
\label{eq: yukawa} 
\end{equation}
would not be appreciably canceled if $\lambda$ is less than the 1~mm thickness of the upper attractor disc.  
We minimized electromagnetic torques by coating the
entire detector with gold and surrounding it by a
gold-coated shield consisting of a tightly-stretched, 10 $\mu$m-thick, beryllium-copper membrane between the detector and attractor plus a copper housing that had small holes for the suspension fiber and the autocollimator beam.  
 The entire system was under a vacuum of $\approx 10^{-6}$ torr in a temperature-controlled and magnetically-shielded environment. The noise in our torque measurements
%, shown in Fig.~\ref{fig: noise}, 
was generally close to the thermal value expected from the finite quality factor, $Q \approx 3000$, of our torsion oscillator, but increased noticeably for detector-membrane separations
below $100~\mu$m (see Fig.~\ref{fig: fft}). A continuous, absolute calibration of the torque scale was provided by the gravitational octupole interaction between 3 small spheres mounted on the detector and 3 larger spheres,
mounted outside the vacuum vessel on a turntable that rotated the spheres about the fiber axis at a steady rate $\omega_{\rm c}$, providing a calibration signal at 3$\omega_{\rm c}$. 

The detector/attractor separation, $\vec{\zeta}=(x,y,s)$,
where $x$ and $y$ are the horizontal displacements between the centers of the detector and attractor and $s$ is the vertical separation between the bottom of the detector ring and the top of the upper attractor disc, was determined using capacitance plus micrometer techniques for $s$, and gravitational plus micrometer techniques  for $x$ and $y$. Detector twist data were taken at $x$ and $y$ close to zero (except for off-center runs used to find the $x,y$ center) and separations $~55\;\mu{\rm m} \leq s \leq 9.53$ mm.
The key parameters, $\eta_n^{\rm exp} \pm \delta \eta_n^{\rm exp}$, of our instrument (the masses removed in machining the holes, the hole radii, thicknesses and locations, etc.) were determined precisely using an electronic balance or a coordinate-measuring machine (CMM).
The CMM readings were corrected for surface-roughness by scanning the relevant surfaces with an  atomic-force-microscope; the correction increased hole diameters and decreased plate thicknesses by 4.4  and 2.3 $\mu$m, respectively. We measured the attractor angle, $\phi$, by counting pulses to the stepper-motor and based $\delta \phi$ on the recorded scatter of the $\phi$ readings when a once-per-revolution index pulse occured.

Our analysis strategy and techniques were described in detail in Ref.~\cite{ho:04}. Briefly, pendulum twist signals were digitally filtered to suppress the free-torsional oscillations and converted into torques by taking into account pendulum inertia and damping as well as signal-averaging and filtering. However, in this work, we made two improvements to our analysis procedure.
\newline
1) We used a more sophisticated 5-point torsion filter that also removed the effects of slow drifts in the equilibrium twist of the torsion pendulum, allowing us to eliminate the polynomial ``drift terms'' that were needed to fit the twist signals in Refs. \cite{ho:01,ho:04}. 
\newline 
2) We mapped the linearity of our autocollimator system after each run by stopping the attractor and calibration turntables and setting 
the pendulum into a free oscillation that covered the same region of our photodetector as the preceeding data run.
\newline
The resulting extracted torques were then decomposed into harmonic amplitudes at multiples of the attractor rotation frequency $\omega$.

The observed harmonic torques, $N_m(\vec{\zeta}_j)\pm \delta N_m(\vec{\zeta}_j)$, where $j$ runs over all $\zeta$ values and $m$=21 or 42 (signals from higher $m$ torques were significantly attenuated by pendulum inertia), were fitted by
predicted torques, $\tilde{N}_m$ that were functions of the key physical parameters of the apparatus, $\vec{\eta}$. We computed the $\tilde{N}_m$ by integrating Eq.~\ref{eq: yukawa} over our measured geometry as described in Ref.~\cite{ho:04}. We first considered the effects of Newtonian gravity alone.
We used gravity to determine independently the key physical parameters of the apparatus, $\vec{\eta}$, and required the two determinations to agree within errors. This was done by minimizing
\begin{equation}
\chi^2=\sum_j \sum_m \left[ \frac{N_m(\vec{\zeta}_j)-\tilde{N}_m(\vec{\zeta}_j,\vec{\eta})}{\Delta N_m (\vec{\zeta}_j)}\right]^2\!\!\!+\sum_n\left[ \frac {\eta_n^{\rm exp}-\eta_n}{\delta\eta_n^{\rm exp}} \right]^2~,
\end{equation}
where the experimental error
\begin{equation}
\Delta N_m (\vec{\zeta}_j)=\sqrt{(\delta N_m(\vec{\zeta}_j))^2+\left(\delta s_j\frac{\partial \tilde{N}_m}{\partial s_j} \right)^2}
\end{equation}
accounted for the uncertainty (typically $\pm 1 \mu$m) in our measurement of the detector's vertical position. This fitting procedure yielded uncertainties that included statistical and most systematic effects, and took into account possible correlations between the uncertainties in the various experimental parameters. Finally, we expanded the analysis to include a single Yukawa interaction.
%
%	Fig 2
%
%
%	Fig 3
%
\begin{figure}
\hfil\scalebox{.53}{\includegraphics*[0.64in,0.5in][6.5in,6.1in]{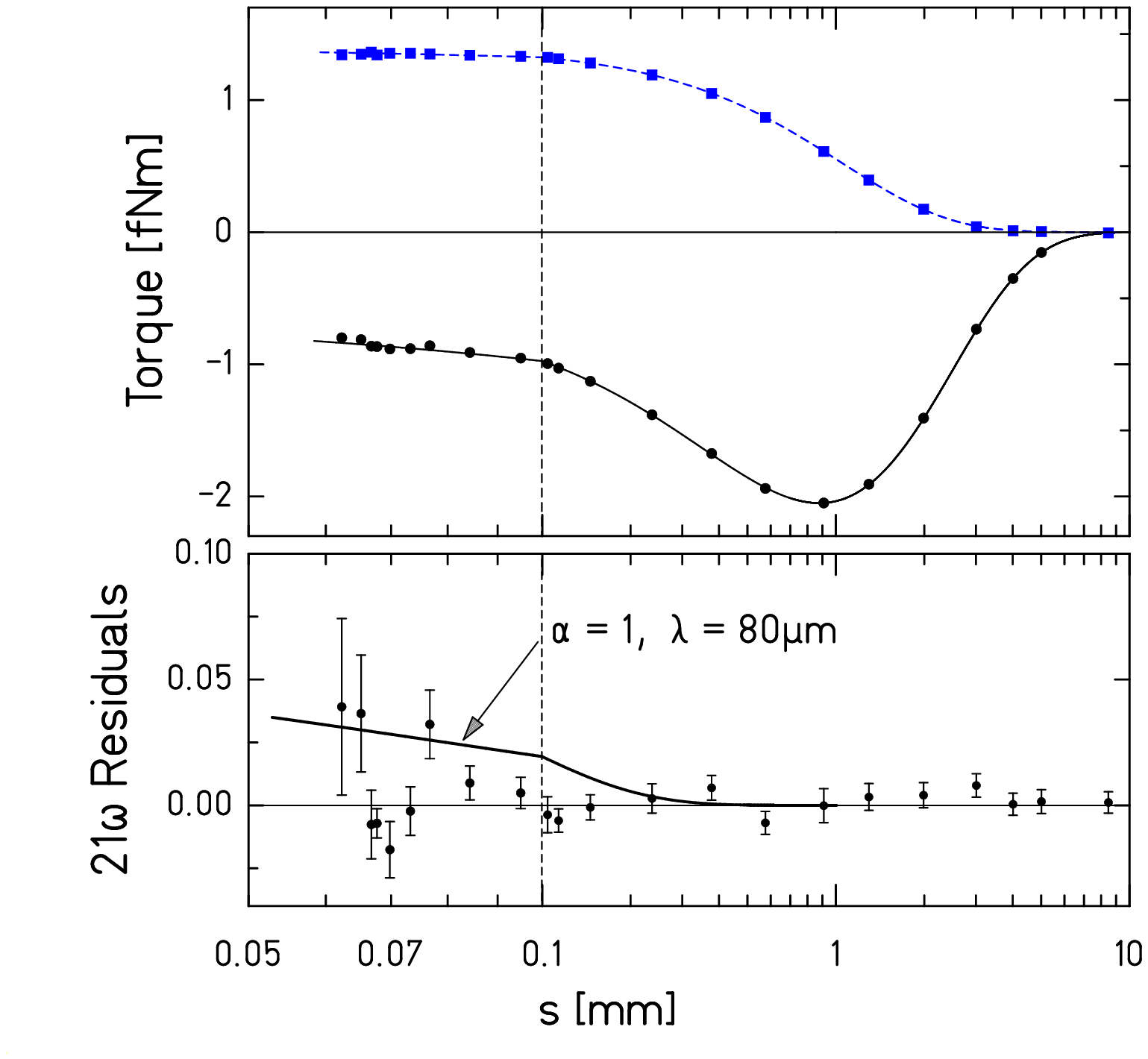}}\hfil
\caption{[Color online] Experiment I torques as functions of vertical separation $s$. The horizontal scale is expanded below $s=100~\mu$m. The upper panel displays the 21$\omega$ and 42$\omega$ torques as solid circles and squares, respectively. 
When not visible the errors are smaller than the size of the points. The smooth curves show the Newtonian fit to the combined data of all 3 experiments.
The lower panel shows the 21$\omega$ residuals; the smooth curve shows the residual that would arise from an $\alpha =1$, $\lambda = 80~\mu$m Yukawa interaction.}
\label{fig: data exp 1}
\end{figure}

We conducted three separate experiments with our apparatus. These used the same detector ring and upper attractor disc, but different lower attractor discs.
In Experiment I the lower-attractor disk thickness was 3.032~mm and the Newtonian 21$\omega$ torque was over-canceled at all values of $s$. In Experiments II and III, this thickness was reduced by 0.140 mm, causing the Newtonian 21$\omega$ torque to be under-canceled for $s<100~\mu$m. On the other hand, a short-range interaction with $\lambda << 1$ mm would produce the same torques in all three experiments, but it must show opposite behaviors in Experiment I compared to Experiments II and III; decreasing, for example, the magnitude of the $21\omega$ signal in one case and increasing it in the other. This feature was useful in discriminating against possible systematic errors. After the two-plate attractor data were taken we accumulated additional data with just the upper or lower attractor plate.

%
%	Fig 4
%
\begin{figure}[t]
\hfil\scalebox{.53}{\includegraphics*[0.64in,0.5in][6.5in,6.1in]{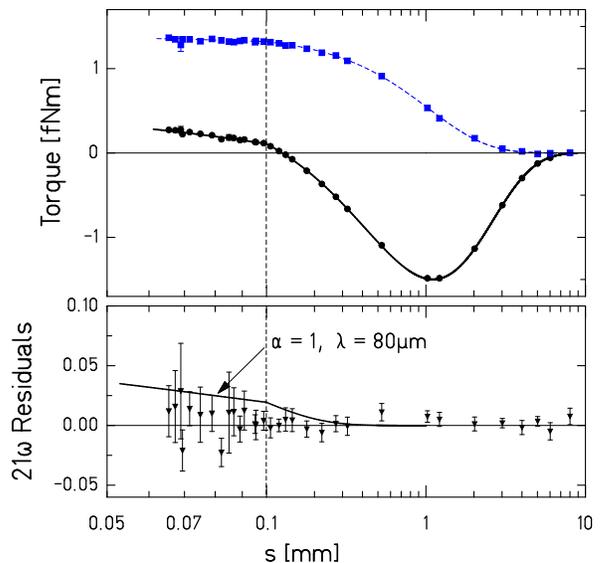}}\hfil
\caption{[Color online] Experiment II torques. Notation is the the same as for Fig. \ref{fig: data exp 1}, but the vertical scales are expanded.} 
\label{fig: data exp 2}
\end{figure}

We used a 20 $\mu$m-diameter suspension fiber in Experiment I, and set the attractor rotation frequency $\omega$ to $\omega_{\rm 0}/28$ and the calibration turntable frequency $\omega_{\rm c}$ to $49\omega/3$, where $\omega_{\rm 0}\approx 12.6$ mrads/s was the detector's free-oscillation frequency. After the Experiment I data were taken we discovered that the detector ring was slightly bowed so that its outer set of holes was slighter higher than the inner set. We accounted for the 
detector's deformation by modeling the outer and inner sets of holes at different average heights, $\Delta z=3.5~\mu$m, above the attractor. (This was not necessary in Ref.~\cite{ho:04} which used a thicker and more rigid detector.)  Data and the Newtonian fit are shown in Fig.~\ref{fig: data exp 1}. A possible deviation at 
$s < 80~\mu$m led us to perform Experiments II and III. 

%
%	Fig 5
%
\begin{figure}[t]
\hfil\scalebox{.53}{\includegraphics*[0.64in,0.5in][6.5in,6.1in]{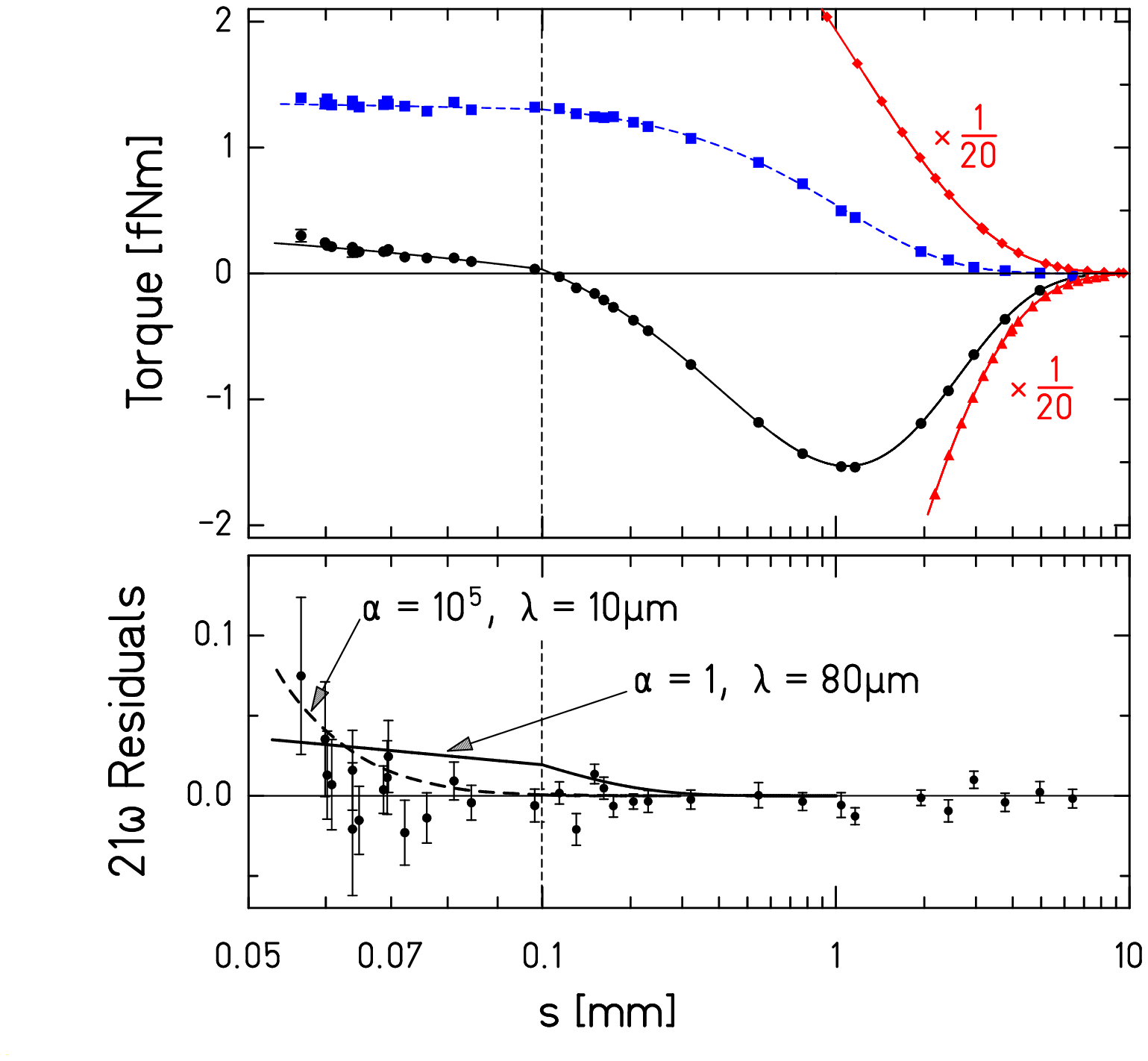}}\hfil
\caption{[Color online] Experiment III torques. Notation is the same as in Fig. \ref{fig: data exp 2}, except that 
diamonds (triangles) show the 21$\omega$ torque from the upper (lower) attractor plate alone. 
The solid and dashed curves in the lower panel show the residuals expected from $\alpha=1$, $\lambda=80~\mu$m and $\alpha=10^5$, $\lambda=10~\mu$m Yukawa interactions, respectively. Both are excluded by our results.}
\label{fig: data exp 3}
\end{figure}

Before beginning Experiment II we flattened the detector ring to $\Delta z \leq 1.0~\mu$m, and switched to a 17 $\mu$m
fiber that gave $\omega_{\rm 0}=9.7$~mrads/s.  We reduced the $1/f$ torque noise from internal losses in the suspension fiber by increasing the attractor and calibration turntable frequencies to $\omega = \omega_{\rm 0}/7.5$ and $\omega_{\rm c}= 6\omega$, respectively. Results from this experiment are shown in Fig.~\ref{fig: data exp 2}.
The small-$s$ anomaly is not evident. Hoping to reduce the excess noise at small $s$, we disassembled the apparatus before Experiment III and replaced the gold coatings on the detector and
beryllium-copper membrane. In the process we inadvertantly increased the deformation of the detector ring to $\Delta z=3.9~\mu$m. We returned to a $20~\mu$m fiber, and operated at $\omega = \omega_{\rm 0}/7.5$  and $\omega_{\rm c}= 3\omega$. Data are shown in Fig.~\ref{fig: data exp 3}. The slightly different  dependence of the  21$\omega$ torque on $s$, compared to Experiment II, is due to the detector curvature in Experiment III. In this case a marginally significant short-range anomaly corresponding to an increased attraction, similar to that seen in Experiment I, is again present. 
%
%	Fig 6
%
\begin{figure}[!]
\hfil\scalebox{.55}{\includegraphics*[64pt,29pt][510pt,431pt]{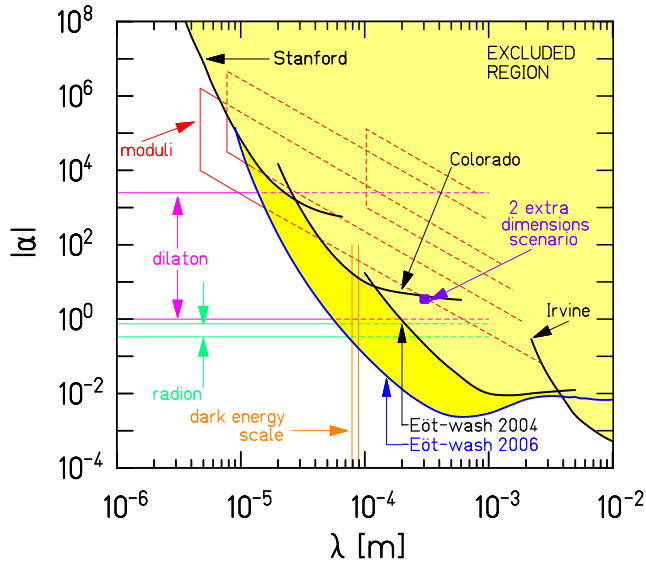}}\hfil
\caption{[Color online] Constraints on Yukawa violations of the gravitational $1/r^2$ law.
The shaded region is excluded at the 95\% confidence level. 
Heavy lines labeled E\"{o}t-Wash 2006, E\"{o}t-Wash 2004, Irvine, Colorado and Stanford show experimental constraints from this work, Refs.~\cite{ho:04}, \cite{ho:85}, \cite{lo:03} and \cite{ch:03,sm:05}, respectively. Lighter lines show various theoretical expectations summarized in Ref.~\cite{ad:03}.}
\label{fig: constraints}
\end{figure}

Non-gravitational backgrounds from magnetic, electrostatic and thermal effects were studied and found to be negligible using techniques described in Ref.~\cite{ho:04}. Casimir forces can only affect our results if they could penetrate the beryllium-copper membrane or cause the membrane to deflect so as to trace the hole pattern of the attractor. Private communications from Robert Jaffe, Paul Chesler, Anton Andreev and Laurence Yaffe have shown that our beryllium-copper membrane was thick enough to reduce direct Casimir forces between the attractor and detector to a negligible level. The observed 1.6 kHz frequency of the lowest ``drum-head'' mode of the membrane was sufficiently high that $m=21$ deformations were insignificant. No corrections for backgrounds were necessary.

A combined Newtonian fit to the data from all 3 experiments gave $\chi^2=407$ for $\nu=421$ degrees of freedom. The best fit with an additional Yukawa interaction improved the $\chi^2$ by 3.5 for $\alpha = -0.0037$, $\lambda = 2~$mm. The combined data showed no evidence for a 2$\sigma$ effect at any $\lambda$.
Our resulting constraints on violations of the inverse-square law, shown in Fig.~\ref{fig: constraints}, improve on previous work by a factor of up to 100. In particular, at 95 \% confidence, we find that any gravitational-strength ($|\alpha|=1$) Yukawa interaction must have $\lambda \leq 56~\mu$m. The results in Fig.~\ref{fig: constraints} yield a model-independent upper limit on the size of a compact extra dimension. A single extra dimension with 
$R \alt s_{\rm min}$ would give a signal corresponding to a Yukawa interaction with $\alpha=8/3$ and $\lambda=R$~\cite{ad:03}, leading to a 95\%-confidence upper bound of $R \leq 44~\mu$m. 
For the two large extra-dimension scenario discussed in Ref.~\cite{ar:98}, we require a $2\sigma$ lower limit on unification mass $M_{\ast} \geq 3.2$ TeV/$c^2$, where $M_{\ast}$ is defined in Ref.~\cite{ho:04}. Constraints from the data in Figs.~\ref{fig: data exp 1},\ref{fig: data exp 2}, \ref{fig: data exp 3}  on other  possible forms of inverse-square-law violation will be submitted as a separate publication.

We thank Anton Andreev, Paul Chesler, Robert Jaffe, Steven Lamoreaux, and Laurence Yaffe for illuminating remarks on the finite-temperature Casimir force.
This work was supported by NSF Grant PHY0355012 and by the DOE Office of Science.

\end{document}